%% file: main.tex
\documentclass[sigconf]{acmart}
\renewcommand\footnotetextcopyrightpermission[1]{}
\AtBeginDocument{%
  }

\setcopyright{acmlicensed}
\copyrightyear{2018}
\acmYear{2018}
\acmDOI{XXXXXXX.XXXXXXX}
\acmConference[Conference acronym 'XX]{Make sure to enter the correct
  conference title from your rights confirmation email}{June 03--05,
  2018}{Woodstock, NY}
\acmISBN{978-1-4503-XXXX-X/2018/06}

\usepackage[utf8]{inputenc} 
\usepackage[T1]{fontenc}    
\usepackage{hyperref}       
\usepackage{url}            
\usepackage{booktabs}       
\usepackage{amsfonts}       
\usepackage{nicefrac}       
\usepackage{microtype}      
\usepackage{xcolor}         
\usepackage{multirow}
\usepackage{multicol}
\usepackage{graphicx}
\usepackage{amsmath}
\usepackage{xspace}
\usepackage{algorithm}
\usepackage{algorithmicx}
\usepackage{algpseudocode}
\usepackage{colortbl} 
\usepackage{tcolorbox}  
\usepackage{framed} 
\usepackage{enumitem}

\newcommand{\ie}{\emph{i.e.,}\xspace}

\begin{document}

\title{Enhancing Sequential Recommendation with World Knowledge from Large Language Models}

\author{Tianjie Dai}
\affiliation{
  \institution{Shanghai Jiao Tong University}
  \country{Shanghai, China}
  }
\email{elfenreigen@sjtu.edu.cn}

\author{Xu Chen$^\dagger$}
\affiliation{
  \institution{Taobao \& Tmall Group}
  \country{Hangzhou, China}
  }
\email{huaisong.cx@taobao.com}

\author{Yunmeng Shu}
\affiliation{
  \institution{Taobao \& Tmall Group}
  \country{Hangzhou, China}
}
\email{yunmeng.shu@hotmail.com}

\author{Jinsong Lan}
\affiliation{%
  \institution{Taobao \& Tmall Group}
  \country{Beijing, China}
}
\email{jinsonglan.ljs@taobao.com}

\author{Xiaoyong Zhu}
\affiliation{%
  \institution{Taobao \& Tmall Group}
  \country{Hangzhou, China}
}
\email{xiaoyong.z@taobao.com}

\author{Jiangchao Yao}
\affiliation{
  \institution{Shanghai Jiao Tong University}
  \country{Shanghai, China}
  }
\email{Sunarker@sjtu.edu.cn}

\author{Bo Zheng}
\affiliation{%
  \institution{Taobao \& Tmall Group}
  \country{Hangzhou, China}
}
\email{bozheng@alibaba-inc.com}

\renewcommand{\shortauthors}{Trovato et al.}

\begin{abstract}
Sequential Recommendation System~(SRS) has become pivotal in modern society, which predicts subsequent actions based on the user's historical behavior. However, traditional collaborative filtering-based sequential recommendation models often lead to suboptimal performance due to the limited information of their collaborative signals. With the rapid development of LLMs, an increasing number of works have incorporated LLMs' world knowledge into sequential recommendation. Although they achieve considerable gains, these approaches typically assume the correctness of LLM-generated results and remain susceptible to noise induced by LLM hallucinations.
To overcome these limitations, we propose \textbf{GRASP}~(\textbf{\underline{G}}eneration Augmented \textbf{\underline{R}}etrieval with Holistic \textbf{\underline{A}}ttention for \textbf{\underline{S}}equential \textbf{\underline{P}}rediction), 
a flexible framework that integrates \textit{generation augmented retrieval} for descriptive synthesis and similarity retrieval, and \textit{holistic attention enhancement} which employs multi-level attention to effectively employ LLM's world knowledge even with hallucinations and better capture users' dynamic interests. The retrieved similar users/items serve as auxiliary contextual information for the later holistic attention enhancement module, effectively mitigating the noisy guidance of supervision-based methods. Comprehensive evaluations on two public benchmarks and one industrial dataset reveal that GRASP consistently achieves state-of-the-art performance when integrated with diverse backbones. 
The code is available at: \url{https://anonymous.4open.science/r/GRASP-SRS}.
\end{abstract}



\maketitle

\vspace{-6pt}
\input{1Introduction}

\input{2RelatedWork}
\input{3Method}
\input{4Experiment}
\input{5Conclusion}

\clearpage
\newpage
\bibliography{main}
\bibliographystyle{abbrv}

\clearpage
\newpage
\appendix
\section{Experimental Setting}
\subsection{Prompt Templates for LLM}
\label{appendix:prompt}
For the construction of prompts for public datasets, we adopted the approach of LLM-ESR~\cite{liu2024llm}. For Industry-100K, since we have more information, we constructed a comprehensive prompt with CoT.

\begin{tcolorbox}[colback=green!5!white, colframe=green!60!black, title=Template for items in Beauty dataset:]
The beauty item has the following attributes: name is <TITLE>; brand is <BRAND>; price is <PRICE>. The item has the following features: <FEATURE>. The item has the following descriptions: <DESCRIPTION>.
\end{tcolorbox}
\begin{tcolorbox}[colback=violet!5!white, colframe=violet!60!black, title=Template for items in Fashion dataset:]
The fashion item has the following attributes: name is <TITLE>; brand is <BRAND>; score is <SCORE>; price is <PRICE>. The item has the following features: <FEATURE>. The item has the following descriptions: <DESCRIPTION>.
\end{tcolorbox}
\begin{tcolorbox}[colback=cyan!5!white, colframe=cyan!60!black, title=Template for users in Beauty and Fashion datasets:]
The user has visited the following items: <TITLE\_1>, <TITLE\_2>, ... please conclude the user's preference.
\end{tcolorbox}
\begin{tcolorbox}[colback=green!5!white, colframe=green!60!black, title=Template for items in Industry-100K dataset:]
You are an e-commerce product analysis expert, skilled in extracting insights from product data and providing comprehensive descriptions of items. Your task is to summarize the product based on its tags, core keywords, personalized recommendation labels, and related categories, paying special attention to modifiers to craft a detailed product introduction.
Rules to follow:

1. The introduction must be thorough and specific, avoiding overly broad statements.

2. It should cover the product's purpose, features, usage scenarios, and commonly paired items.

3. If there is insufficient information to answer, output ``NA'' instead of fabricating details.

Task steps:

1. First, gain a basic understanding of the product based on the provided product tags, core keywords, personalized recommendation labels, and related categories.

2. Next, derive details about its purpose, features, usage scenarios, and commonly paired items from this understanding.

3. Finally, generate a structured output based on the extracted information, ensuring no repetition.

Prioritize accuracy, but also think divergently to ensure both correctness and richness in the output.

Product information: Product tag: <TITLE>, core keyword: <CORE\_WORD>, personalized recommendation tags: <TAGS>, related category info: <COMMODITY\_INFO>.
\end{tcolorbox}

\begin{tcolorbox}[colback=cyan!5!white, colframe=cyan!60!black, title=Template for users in Industry-100K dataset:]
You are an expert in e-commerce user behavior analysis and skilled in summarizing user profiles based on their e-commerce activities. The user's birthplace is <BIRTH\_CITY>, gender is <GENDER>, age is <AGE>, zodiac sign is <CONSTELLATION>, current residence is <PROV\_NAME><CITY\_NAME>, mobile phone brand is <PHONE\_BRAND>, graduated from <PRED\_SCHOOL>, occupation type is <PRED\_CAREER\_TYPE>, education level is <PRED\_EDUCATION\_DEGREE>, marital status is <PRED\_LIFE\_STAGE\_MARRIED>, user spending power level is <USER\_PP\_LEVEL>, user activity level is <ACTIVITY>, and recent purchase history is <HIST\_INFO\_PROMPT>. Please analyze the user's personalized preferences according to the following steps and provide a summary:

1. Analyze the user's regional online shopping behavior preferences based on their personalized characteristics. Assess potential purchase preferences based on their occupation, education level, and marital status. Evaluate their spending power and activity level based on attributes such as mobile phone brand, housing price, and spending power level.

2. Analyze the user's recent product content preferences and acceptable price range based on the titles, categories, and prices of recently purchased items.

3. Understand the temporal evolution of the user's preferences by examining the sequence of their recent purchases.

4. Compare the user's personalized characteristic analysis (Step 1) with their purchasing behavior analysis (Steps 2 and 3), then provide user profile labels and recommendation directions.
\end{tcolorbox}

\subsection{Dataset Statistics}
\label{appendix:dataset}
In this paper, comprehensive experiments are carried out on two widely utilized public datasets and one industrial dataset. The Beauty dataset, sourced from Amazon\footnote{\url{https://cseweb.ucsd.edu/~jmcauley/datasets.html\#amazon_reviews}}~\cite{mcauley2015image}, consists of user reviews on beauty-related products. Similarly, the Fashion dataset, also part of the Amazon collection, contains reviews on fashion items. The Industry-100K dataset is a subset curated from user purchase records on an e-commerce platform, capturing transactions spanning from January 17, 2025, to February 23, 2025. It encompasses data from approximately 100,000 users, providing a comprehensive snapshot of purchasing behaviors during this period. 

For preprocessing, we follow SASRec~\cite{kang2018self} and LLM-ESR~\cite{liu2024llm}. For data partitioning, we adopt the \textit{leave-one-out} manner. Furthermore, we partition the data into \textit{head} and \textit{tail}. For the Beauty dataset, the demarcation point for head/tail users is set at 9, and for items at 4. In the case of the Fashion dataset, the threshold for users is 3, and for items it is 4. As for the Industry-100K dataset, the criterion for users is 29, and for items it is 2. 

\subsection{Theoretical Analysis}

The core difference between GRASP and existing works (e.g., LLM-ESR) lies in their robustness to LLM hallucination, which is evident from a gradient perspective.

\textbf{Existing works} (e.g., LLM-ESR) use the LLM-generated context as a rigid supervisory signal via a regularization term: $\mathcal{L} = \mathcal{L}_{\text{main}} + \lambda \lVert \mathbf{e}_u - \mathbf{e}_c \rVert_2^2$, where $\mathbf{e}_u$ denotes the user embedding and $\mathbf{e}_c$ denotes the context embedding derived from the LLM. The gradient of this regularization term includes a component $2(\mathbf{e}_u - \mathbf{e}_c)$, which unconditionally pulls $\mathbf{e}_u$ toward $\mathbf{e}_c$, even when the LLM context is inaccurate or hallucinated.

\textbf{GRASP}, in contrast, treats the LLM context as an input feature that is fused with the user embedding through a learnable transformation: $\mathbf{h}_u = \mathbf{W}_e \mathbf{e}_u + \mathbf{W}_c \mathbf{e}_c + \mathbf{b}$, where $\mathbf{W}_e$, $\mathbf{W}_c$, and $\mathbf{b}$ are learnable parameters. This formulation effectively implements a learnable gating mechanism. Crucially, the gradient for the context weight $\mathbf{W}_c$ is driven solely by the main task loss $\mathcal{L}_{\text{main}}$. If the LLM context is noisy or uninformative, the model can learn to down-weight it---i.e., $\mathbf{W}_c \to \mathbf{0}$---to minimize $\mathcal{L}_{\text{main}}$. This design adaptively filters out hallucinated content based on its utility for the downstream task.

\subsection{Online Production Validation}
We trained GRASP on a larger-scale internal industry dataset containing 10 million users and integrated its learned user representations as input features into a downstream CTR prediction model. Using 32 A100 GPUs for training and 16 for daily updated user representations inference, we show that our enhanced model provides useful information over the online CTR baseline by an online A/B test conducted in Aug 2025 on our e-commerce platform with 50 million daily active users, using 5\% traffic allocation. The results show a 0.14 point absolute increase in CTR, a 1.69\% relative growth in order volume, and a 1.71\% uplift in GMV, confirming GRASP's practical value in real-world applications.




\end{document}

%% file: 1Introduction.tex
\section{Introduction} \label{sec:intro}

\footnotetext[1]{$\dagger$ Corresponding author.}

\begin{figure*}[t]
\centerline{\includegraphics[width=\linewidth]{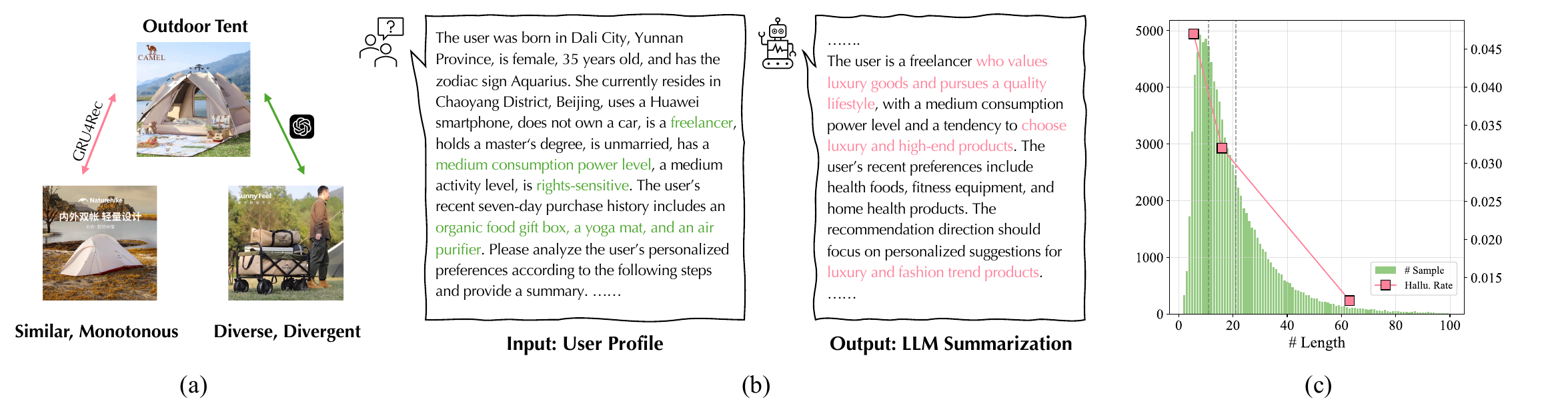}}
\caption{(a) GRU4Rec over-emphasizes frequent interactions while underrepresenting diverse or contextually related user intents, which was trained on the Industry-100K dataset collected from an e-commerce platform. (b) An example demonstrating LLM hallucinations, where generated content deviates from reality. LLM erroneously categorizes the consumption habits of this user with moderate purchasing power and a focus on health and wellness as opting for luxury goods. (c) Sequence length distribution of Industry-100K and hallucination rate analysis across different groups based on the user's interaction sequence length, where we can observe a significant increase in hallucinations as sequence length decreases.}
\label{fig:intro-motivation}
\end{figure*}

Sequential Recommendation System~(SRS) has emerged as a critical component in modern recommendation scenarios, which leverages sequential user-item interactions to model users' temporal patterns and predict future behaviors~\cite{kang2018self,xie2022contrastive,chen2018sequential,boka2024survey,yang2023debiased}. By capturing the dynamic nature of user preferences, SRS enables personalized recommendations across diverse domains, including e-commerce, short-video platforms, and social networks~\cite{nasir2023survey,zheng2024large,gao2022kuairand}. Despite their success, the performance of SRS models still has considerable room for improvement, as they are predominantly built upon ID embeddings that neglect much other valuable information.~\cite{chen2025enhancing,pan2024survey,yuan2023go}

The fundamental limitations of ID-based collaborative filtering approaches stem from their inherent inability to capture comprehensive user behavior patterns. As user behavior in real-world scenarios is often scattered and multi-faceted, relying solely on collaborative signals inevitably introduces bias. As illustrated in Fig.~\ref{fig:intro-motivation}(a), traditional ID-based models tend to prioritize frequently interacted items while overlooking more diverse and contextually broader user intentions. For instance, when a user purchases an outdoor tent, their underlying interest may extend to a wider camping ecosystem—including sleeping bags, portable stoves, and lighting equipment. However, models like GRU4Rec predominantly focus on items that are highly similar to previously interacted ones, like recommending various types of tents. This overemphasis on dominant signals leads to a failure in capturing users’ latent, cross-category interests. Consequently, purely collaborative modeling of sequential behavior often leads to suboptimal solutions~\cite{yuan2023go,li2025id,jiang2024reformulating}.

Recent advancements in large language models have opened new avenues for enhancing SRS by generating rich semantic descriptions of users and items with the integration of world knowledge, which provides more diverse and comprehensive signals rather than collaborative information~\cite{li2024calrec,harte2023leveraging,boz2024improving,hu2024enhancing}. For instance, LLM-ESR~\cite{liu2024llm} leverages the world knowledge understanding capabilities of LLMs to generate semantic features, which are then used to identify similar users for supervision. Specifically, LLM-ESR designs a loss function to ensure that the behavioral representations of these semantically similar users also exhibit similarity to enhance recommendation performance. While promising, this approach still has some drawbacks. As shown in Fig.\ref{fig:intro-motivation}(b), LLMs are prone to hallucination issues, where they may generate descriptions that deviate from reality or contain inaccuracies~\cite{huang2025survey,rawte2023survey}. Furthermore, we observe that when using LLM to generate user descriptions with user histories, the probability of hallucinations varies for users with different interaction lengths. As illustrated in Fig.~\ref{fig:intro-motivation}(c), we evaluate hallucination rates on a real-world industrial dataset, Industry-100K, collected from an internal e-commerce platform. We randomly sample 500 users for three user groups of different history length, and generate their semantic descriptions using Qwen2.5-7B-Instruct~\cite{yang2024qwen2}. Through human evaluation, we find that hallucination rates increase significantly as the length of user interaction sequences decreases. Consequently, directly using these potential hallucinated semantic features as supervision signals risks introducing noise, which could significantly impair the model's performance and reliability. \textit{Therefore, how to better leverage the generated world knowledge from LLMs to supplement information in SRS, is a critical issue that needs further studying.}

In this paper, we design \textbf{GRASP}, a novel framework that integrates LLM-derived world knowledge with traditional sequential recommendation models. It leverages the understanding capabilities of LLMs to generate semantic representations and dynamically fuses multi-perspective signals, which can not only supplement the sparse collaborative signals but also circumvent the potential erroneous semantic embeddings brought by hallucination. GRASP boosts SRS through two-fold components, \ie \textit{generation augmented retrieval} and \textit{holistic attention enhancement}. On one hand, \textit{generation augmented retrieval} employs LLMs to generate detailed user profiles and item descriptions, constructing an offline database for retrieval. By leveraging semantic embeddings, it identifies top-$k$ similar users or items to provide auxiliary information in a semantic, low-dimensional space, rather than relying on sparse, high-dimensional representations. This approach effectively integrates the original attributes with the LLM's world knowledge, projecting them into a compact embedding space to facilitate efficient similarity-based retrieval. On the other hand, \textit{holistic attention enhancement} integrates retrieved information as contextual input rather than direct supervision, effectively avoiding noisy guidance from potential LLM hallucinations. Specifically, it employs a multi-level attention mechanism: initial user-item attention captures core interaction patterns, followed by attention between similar user/item groups to incorporate neighborhood context, and finally concatenated attention for global interest modeling. Extensive experiments on three benchmarks demonstrate GRASP's significant improvements on multiple SRS baselines in the overall recommendation performance. Our contributions can be summarized as follows:
\begin{itemize}[leftmargin=*]
    \item We propose GRASP, a flexible framework designed orthogonal to current sequential recommendation systems, while simultaneously mitigating the inaccuracies associated with attribute-based retrieval and preventing noisy guidance issues arising from the direct utilization of LLM-generated hallucinated content as supervision signals.

    \item We introduce \textit{generation augmented retrieval} and \textit{holistic attention enhancement}, to leverage information from similar users/items as supplement information while effectively capturing dynamic user interests.
    
    \item Empirically, GRASP outperforms state-of-the-art baselines on two public datasets and one industry benchmark, exhibiting consistent advantages in overall recommendation performances under diverse settings. 

\end{itemize}

%% file: 2RelatedWork.tex
\begin{figure*}[t!]
\centerline{\includegraphics[width=\linewidth]{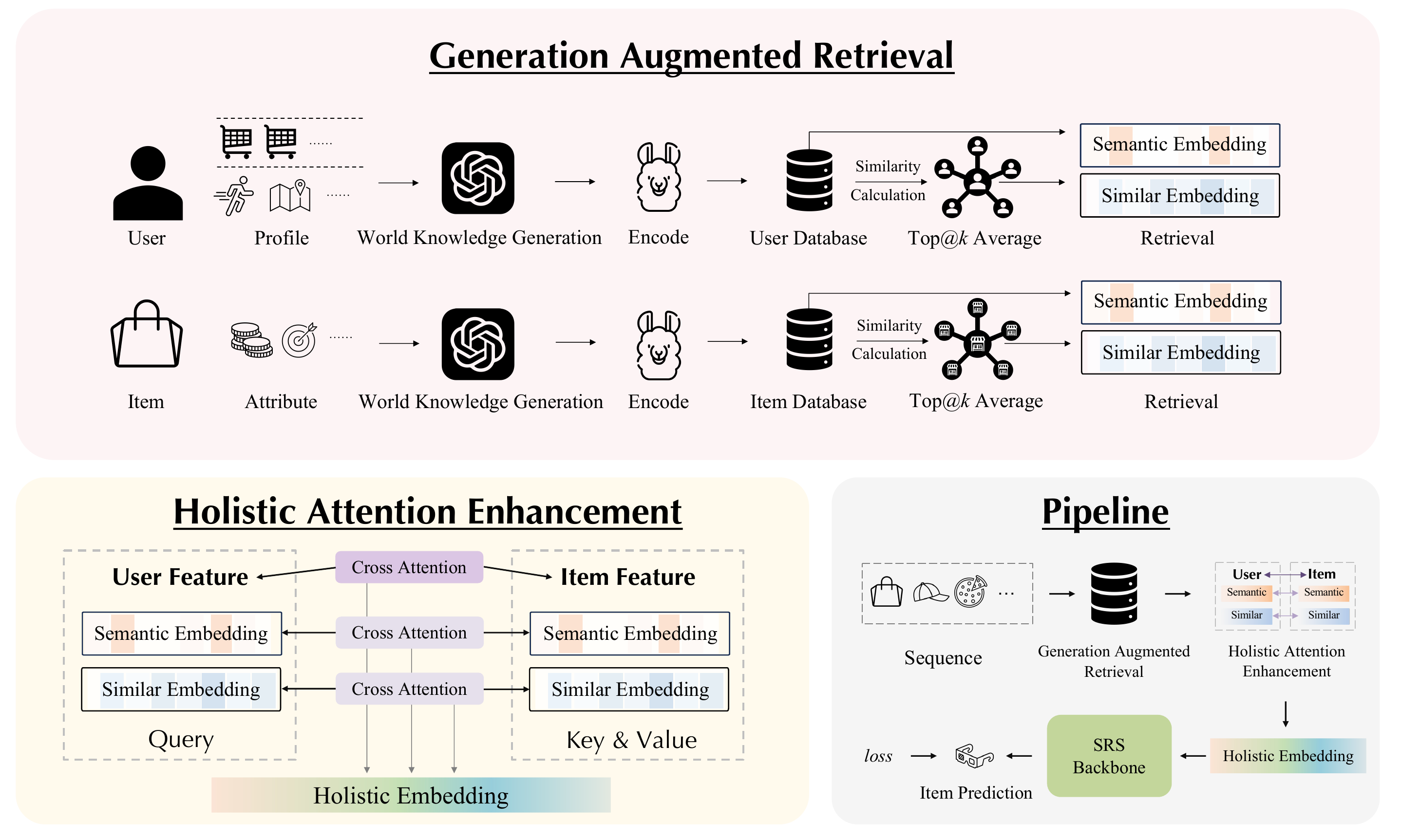}}
\caption{The overview of GRASP framework, which is flexible on top of sequential recommendation baselines, and consists of generation augmented retrieval and holistic attention enhancement. The pipeline demonstrates the workflow and roles of each module within GRASP.
}
\label{fig:framework}
\end{figure*}

\section{Related Works}
\subsection{Sequential Recommendation System}
The goal of sequential recommendation models is to predict the content that a user is likely to interact with at the next moment based on their historical interaction behaviors~\cite{quadrana2018sequence,mishra2015web,wang2019sequential,ying2018sequential,boka2024survey,chang2021sequential,liu2016context,tang2018personalized,xu2019recurrent}. GRU4Rec, as one of the pioneering works in SRS, leverages GRU to capture sequential patterns in user behavior sequences~\cite{hidasi2015session}. SASRec employs a self-attention mechanism to capture long-range dependencies and complex item relationships without sequential computation constraints~\cite{kang2018self}. Bert4Rec leverages the masked language modeling paradigm to learn contextual representations through masked item prediction pre-training task~\cite{sun2019bert4rec}. However, these conventional SRSs typically rely on collaborative filtering representations for users or items, neglecting other valuable signals from multiple perspectives.

\vspace{-2pt}
\subsection{LLM for Sequential Recommendation}
Recent years have seen LLMs excel in natural domains, with growing research integrating them into search and recommendation systems~\cite{acharya2023llm,kim2024large,fan2024recommender,liu2025llmemb,jia2025learn,wang2024towards,gao2024llm,wang2024llm4msr,zhang2025does,feng2025iranker,zheng2024adapting}. Broadly speaking, these works can be categorized into two paradigms. The first paradigm involves using LLMs as the entire system. For example, ProLLM4Rec~\cite{xu2024prompting} designs prompts to leverage the capabilities of LLMs in recommendation tasks. TALLRec~\cite{bao2023tallrec} creates an instruction-tuning dataset from user history, fine-tunes LLM, and uses its outputs directly as recommendation predictions. However, this paradigm relies heavily on LLMs during inference, leading to high computational costs and difficulty meeting industrial demands for high concurrency and low latency. The second paradigm focuses on integrating the semantic understanding capabilities of LLMs into recommendation systems. For instance, HLLM~\cite{HLLM} designs item LLM  to extract rich content features from text description of the item, and user LLM to utilize these features to predict users’ future interests. LLM-ESR~\cite{liu2024llm} utilizes the hidden embedding of LLMs to initialize ID embeddings and designs a dual network along with a self-distillation loss function that leverages LLM embeddings to search for similar users, thereby enhancing the performance of various traditional sequential recommendation models. LRD~\cite{yang2024sequential} leverages LLMs to discover latent item relations through language representations and integrates them with a discrete state variational autoencoder to enhance relation-aware sequential recommendation. \textit{Unlike methods that directly use LLM embeddings as supervision signals which may amplify noise due to potential hallucination, our approach leverages LLM's semantic understanding capability and integrates similar users/items as auxiliary information to improve model performance.}

%% file: 3Method.tex
\vspace{-8pt}

\section{Methodology}
In this section, we first provide a brief introduction to the sequential recommendation model, followed by a detailed description of our proposed method, \textbf{GRASP}, which consists of generation augmented retrieval and holistic attention enhancement. The framework of GRASP is shown in Fig.~\ref{fig:framework}.

\subsection{Problem Formulation}
Let $\mathcal{S}_u = \{i_1, i_2, \dots\}$ represents the interaction sequence of user $u$, where $i_j$ denotes the $j$-th item interacted with the user. Here, $u \in \mathcal{U} = \{u_1, u_2, \dots, u_n\}$ is a user from the set of $n$ users, and $\mathcal{I} = \{i_1, i_2, \dots, i_m\}$ is the item set. The objective of the SRS is to predict the next item a user is most likely to interact with. Mathematically, this task can be formulated as:
\begin{equation}
\label{eq:formulation}
    i^* = \underset{i_j \in \mathcal{I}}{\text{argmax}} \; f(i_{|\mathcal{S}_u|+1} = i_j \mid \mathcal{S}_u)
\end{equation}
where $|\mathcal{S}_u|$ denotes the user's historical sequence length, and $f$ is the SRS which outputs the probability of $i_j$ being the next interaction given $\mathcal{S}_u$. \textit{Our method primarily focuses on enhancing the representation of $i$ and $u$, while remaining orthogonal to the SRS backbone.}

\subsection{Generation Augmented Retrieval}
Unlike classical sequential recommendation models that directly initialize the embedding matrix as representations of items or users, we aim to leverage the semantic understanding capabilities of large language models to enrich the semantic information of embeddings and strengthen the modeling of users and items. To achieve this goal, we design a Generation Augmented Retrieval paradigm.

\textbf{Generation.} 
We first construct prompt templates~(detailed in the Appendix~\ref{appendix:prompt}) that incorporate item attributes or user profiles and historical behaviors, respectively. The LLM is then invoked to interpret item information and user preferences, generating descriptive text for all items and users. Next, we extract embeddings from the generated text to facilitate subsequent model operations. For open-source datasets, embeddings are directly obtained from the API\footnote{Following LLM-ESR~\cite{liu2024llm}, we adopt \texttt{text-embedding-ada-002} via OpenAI's API~\url{https://platform.openai.com/docs/guides/embeddings}.}; for internal industrial data, we employ an open-source text encoder~(detailed in Section~\ref{sec:exp}) to extract. As a result, we construct two semantic embedding databases: $\mathbf{U}$ for users and $\mathbf{I}$ for items. 

\textbf{Retrieval.} 
To enhance feature representation, particularly to supplement feature information in scenarios with sparse interaction data, we adopt a nearest-neighbor retrieval strategy. For each user/item, we retrieve the top-$k$ most similar users/items based on the cosine similarity between embeddings. These embeddings are then aggregated via average pooling. Formally, for a given user $u$ and item $i$, the retrieval process can be expressed as:
\begin{equation}
\begin{split}
    \bar{\mathbf{u}} &= \text{Avg\_Pooling}(\mathbf{u}_i \mid \mathbf{u}_i \in \text{Top@k}(\mathbf{u}) \setminus \{\mathbf{u}\}) \\
    \bar{\mathbf{i}} &= \text{Avg\_Pooling}(\mathbf{i}_j \mid \mathbf{i}_j \in \text{Top@k}(\mathbf{i}) \setminus \{\mathbf{i}\})
    \label{eq:retrieval}
\end{split}
\end{equation}
Here, $\bar{\mathbf{u}}$ and $\bar{\mathbf{i}}$ represent the averaged embeddings of the top-$k$ similar users/items. Thus, each user/item is not only represented by its original embedding but also enriched with embeddings from its nearest neighbors. In subsequent stages, we \textit{freeze} and \textit{cache} these embeddings.

\subsection{Holistic Attention Enhancement}

In the previous step, we obtain LLM database for all user preferences and item features, denoted as $\mathbf{U} \in \mathbb{R}^{n \times d}$ and $\mathbf{I} \in \mathbb{R}^{m \times d}$, where $n$ and $m$ represent the number of users and items, respectively, and $d$ is the semantic embedding dimension. Specifically, for each user $u_i$, we have their LLM embedding $\mathbf{u}_i \in \mathbb{R}^d$ as well as the averaged embeddings of corresponding similar users $\bar{\mathbf{u}}_i \in \mathbb{R}^d$, and the same applies to items $i_j$, with $\mathbf{i}_j \in \mathbb{R}^d$ and $\bar{\mathbf{i}}_j \in \mathbb{R}^d$.

To grasp the user's dynamic interests along their historical item sequences, we uniformly treat the user embedding as the query $\mathbf{q}$ and the item embedding as both the key and the value, denoted by $\mathbf{v}$ (since the key and value are identical). We perform an attention-based fusion operation, where the attention mechanism is defined as follows:
\begin{equation}
{\mathcal{A}}(\mathbf{q}, \mathbf{v}) = \sigma\left(\frac{\mathbf{q} \mathbf{v}^T}{\sqrt{d}}\right) \mathbf{v}
\end{equation}
Here, $\sigma$ denotes \textit{Sigmoid} function, which replaces the traditional softmax function. This choice avoids the single-peak issue inherent in softmax, allowing for a representation that better reflects users' diverse preferences while maintaining more raw interest patterns.

Given the user semantic embedding $\mathbf{u}_i$, item semantic embedding $\mathbf{i}_j$, and their corresponding averaged similar embeddings $\bar{\mathbf{u}_i}$ and $\bar{\mathbf{i}_j}$, the holistic attention-enhanced embeddings are computed through a series of attention operations. To enrich the input information, we concatenate the user embeddings with the averaged embeddings of similar users, and the item embeddings with the averaged embeddings of similar items to form global embeddings. Specifically, the holistic attention-enhanced embeddings are calculated as follows:
\begin{equation}
\begin{gathered}
    \mathbf{i}_{j,\text{self}}^{\text{HAE}} = \mathcal{A}(\mathbf{u}_i, \mathbf{i}_j), \quad \mathbf{i}_{j,\text{similar}}^{\text{HAE}} = \mathcal{A}(\bar{\mathbf{u}}_i, \bar{\mathbf{i}}_j) \\
    \mathbf{i}_{j,\text{global}}^{\text{HAE}} = \mathcal{A}([\mathbf{u}_i \parallel \bar{\mathbf{u}}_i], [\mathbf{i}_j \parallel \bar{\mathbf{i}}_j])
    \label{eq:enhanced_cal}
\end{gathered}
\end{equation}
\textit{The fine-grained attention, together with the global attention operation, ensures that the embeddings capture both individual characteristics and aggregated patterns from users and items, leading to a multi-level and comprehensive representation.} These enhanced vectors are then concatenated and passed through MLP to fit the input size of the SRS backbone $f$:
\begin{equation}
\label{eq:enhanced_input}
\mathbf{i}_{j,all} = \text{MLP}\left([\mathbf{i}_{j,self}^{\text{HAE}} \parallel \mathbf{i}_{j,similar}^{\text{HAE}} \parallel \mathbf{i}_{j,global}^{\text{HAE}}\right])
\end{equation}
By performing attention operations directly on the LLM embeddings before mapping them to the hidden dimension of the SRS model, we ensure that semantic information remains intact. Additionally, the inclusion of similar users and items as auxiliary inputs enhances overall information richness of the representation, providing a robust foundation for training and inference in the SRS backbone $f$.

\subsection{Training and Deployment Complexity}
As previously mentioned, our approach primarily focuses on the embedding enhancement of the model, leveraging similar users/items as auxiliary information with a multi-level holistic attention mechanism rather than supervised signals. Consequently, our method can be flexibly integrated on top of existing SRS backbones. 

The overall training objective is the standard loss function used in the SRS backbone~\cite{hidasi2015session,sun2019bert4rec,kang2018self}, as defined in Eq.~\eqref{eq:main}. Here, $y \in \{0, 1\}^{|\mathcal{B}|}$ denotes the ground truth of items in candidate pool $\mathcal{B}$, $\hat{y}_j$ is the predicted probability for item $j$,  $\mathbf{o}$ represents the user representation learned by SRS, and $\mathbf{i}_{j,all}$ is the holistic item embedding computed via Eq.~\eqref{eq:enhanced_input}.
\begin{equation}
\begin{split}
    \mathcal{L} &= -\frac{1}{\mathcal{|B|}}\sum_{j} \left[
        y_j \log(\hat{y}_j) + (1 - y_j) \log(1 - \hat{y}_j)
        \right], \\
    \hat{y}_j &= \sigma\left(\mathbf{o} \cdot \mathbf{i}_{j,all}\right)
    \label{eq:main}
\end{split}
\end{equation}
\textit{For practical deployment, since the LLM-generated embeddings are offline precomputed daily along with the user behavior changes in each day, our method does not introduce excessive online computational overhead.} The increased online computation is the holistic attention module. This module has limited time complexity $\mathcal{O}(l^2d)$ when the sequence length $l$ and latent dimension $d$ in following SRS backbone are fixed. Notice that the similar users and items can also be offline pre-retrieved daily in a small group rather than the whole users and items for efficient industrial deployment. For example, given a user/item, we only retrieve its similar neighbor users/items in the same group or category, largely alleviating the nearest-neighbor search complexity in deployment.

\begin{algorithm}[tbp]
\caption{Pseudo code of GRASP.}
\label{alg}
\begin{algorithmic}[1]
\Require Interaction sequence $\mathcal{S}_u$.
\State Generate LLM embedding database $\mathbf{U}$, $\mathbf{I}$; retrieve similar user/item and generate $\bar{\mathbf{U}}$, $\bar{\mathbf{I}}$ by Eq.~\eqref{eq:retrieval}.
\Statex \textbf{\textit{Training}}
\State Freeze $\mathbf{U}$, $\mathbf{I}$, $\bar{\mathbf{U}}$ and $\bar{\mathbf{I}}$.
\For{each iteration}
\State Compute fine-grained and global enhanced embedding using Eq.~\eqref{eq:enhanced_cal}.
\State Compute input sequence embedding $\mathbf{i}_{all}$ after holistic attention by Eq.~\eqref{eq:enhanced_input}.
\State Calculate loss function $\mathcal{L}$ using Eq.~\eqref{eq:main}.
\State Update model parameters.
\EndFor
\State \textbf{Return}
\Statex \textbf{ }

\Statex \textbf{\textit{Testing}}

\For {$u$ in $\mathcal{U}$}
        \State Obtain corresponding input embedding from $\mathbf{U}$, $\mathbf{I}$, $\bar{\mathbf{U}}$ and $\bar{\mathbf{I}}$, obtain the model parameters.
        \State Compute the scores of items in the candidate set by Eq.~\eqref{eq:formulation} and return the ranked order.
\EndFor

\end{algorithmic}

\end{algorithm}

%% file: 4Experiment.tex
\section{Experiments}
\label{sec:exp}
\subsection{Experimental Setup}
\textbf{Benchmarks}.
We conduct experiments on two publicly available datasets and one collected industry dataset. The two public datasets are Amazon Beauty and Fashion~\cite{mcauley2015image}. The Industry-100K dataset is one subset which was collected from user purchase records spanning from January 17, 2025, to February 23, 2025 on an e-commerce platform, comprising approximately 100,000 users. We follow the previous sequential recommendation works~\cite{liu2024llm} for data preprocessing. The statistical descriptions of the dataset are shown in Table~\ref{tab:exp_dataset}.
\vspace{-3pt}
\begin{table}[h]
\centering
\small
\caption{Statistics of datasets.}
\begin{tabular}{ccccc}
\toprule[1pt]
\textbf{Dataset} & \textbf{\# User} & \textbf{\# Item} & \textbf{\# AVG Length} & \textbf{Sparsity} \\
\midrule
Beauty & 52204 & 57289 & 7.56 & 99.99\%\\
Fashion & 9049 & 4722 & 3.82 & 99.92\% \\
Industry-100K & 99711 & 1205282 & 20.88 & 99.99\%\\
\bottomrule[1pt]
\end{tabular}
\label{tab:exp_dataset}
\end{table}

\textbf{Baselines}.
Since our method is orthogonal to different backbones, we combine our methods on three classic baseline models for sequential recommendation, \ie GRU4Rec~\cite{hidasi2015session}, BERT4Rec~\cite{sun2019bert4rec} and SASRec~\cite{kang2018self}. We compare our method with other LLM-enhanced sequential recommendation models like RLMRec~\cite{ren2024representation}, LLMInit~\cite{harte2023leveraging,hu2024enhancing}, and LLM-ESR~\cite{liu2024llm}.

\textbf{Evaluation Metrics}.
To comprehensively assess the performance of the models, we utilize Normalized Discounted Cumulative Gain~(NDCG) and Hit Rate~(HR) as evaluation metrics, with ranking positions \( k \in \{1, 3, 5, 10, 20\} \). In terms of data split, we adopt \textit{leave-one-out} manner for validation and testing. During the evaluation, the size of the negative sampling is 100. 

\textbf{Implementation Details}.
All experiments are conducted on a single NVIDIA A100 GPU. To ensure fairness, we set the maximum sequence length to 100 and fix the hidden embedding dimension to 64 for all methods. The batch size is 128, and the Adam optimizer is employed with a fixed learning rate of 0.001. Early stopping is applied if NDCG@10 on the validation set does not improve over 20 consecutive epochs. To ensure robustness, we report average results from three tests with random seeds $\{42, 43, 44\}$. For the public datasets, we utilize OpenAI API to obtain LLM embeddings~\cite{liu2024llm}, with an embedding dimension of 1536. For the industry dataset, due to data confidentiality requirements, we employ Qwen2.5-7B-Instruct~\cite{yang2024qwen2} to generate descriptive texts and utilize the pre-trained text encoder LLM2Vec~\cite{behnamghader2024llm2vec} to acquire semantic embeddings, with a dimension of 4096.

\begin{table*}[t!]
\centering
\caption{Overall performance comparison of different methods on three datasets. (\textbf{Bold} indicates the best performance; \underline{Underline} indicates the second best performance; N = NDCG, H = Hit Rate; We report the results of the two-sided t-test with $p < 0.05$.)}
\label{tab:main_exp}
\resizebox{1\textwidth}{!}{
\footnotesize
\begin{tabular}{cc|ccccc|ccccc}
\toprule[1pt]

\textbf{Dataset} & \textbf{Model} & \textbf{N@1} & \textbf{N@3} & \textbf{N@5} & \textbf{N@10} & \textbf{N@20} & \textbf{H@1} & \textbf{H@3} & \textbf{H@5} & \textbf{H@10} & \textbf{H@20} \\ \cmidrule{1-12}

\multirow{15}{*}{\textbf{Beauty}} 
& GRU4Rec &  11.48	& 16.88	& 19.23	& 22.42	& 25.62	& 11.48	& 20.83	& 26.56	& 36.45	& 49.17 \\
& - RLMRec & 11.03&16.50&18.93&22.25&25.48&11.03&20.51&26.41&36.75&49.60\\  
& - LLMInit & 14.33 & 20.53 & 23.05 & 26.29 & 29.37 & 14.33 & 25.08 & 31.20 & 41.24 & 53.46 \\  
& - LLM-ESR & \underline{17.50} & \underline{25.25} & \underline{28.18} & \underline{31.82} & \underline{35.05} & \underline{17.50} & \underline{30.88} & \underline{38.02} & \underline{49.28} & \underline{62.09} \\
& \cellcolor{gray!10}- \textbf{GRASP} & \cellcolor{gray!10}{\textbf{18.15}}&  
\cellcolor{gray!10}{\textbf{26.88}}&  
\cellcolor{gray!10}{\textbf{30.21}}&  
\cellcolor{gray!10}{\textbf{34.16}}&  
\cellcolor{gray!10}{\textbf{37.56}}&  
\cellcolor{gray!10}{\textbf{18.15}}&  
\cellcolor{gray!10}{\textbf{33.24}}&  
\cellcolor{gray!10}{\textbf{41.35}}&  
\cellcolor{gray!10}{\textbf{53.57}}&  
\cellcolor{gray!10}{\textbf{67.03}} \\
\cmidrule{2-12} 
& BERT4Rec & 10.49 & 16.86 & 19.68 & 23.33 & 26.67 & 10.49 & 21.56 & 28.43 & 39.72 & 52.93\\
& - RLMRec & 10.45&16.82&19.66&23.20&26.72&10.45&21.50&28.42&39.43&53.39\\  
& - LLMInit & 16.57 & 24.74 & 27.92 & 31.65 & 34.89 & 16.57 & 30.71 & 38.45 & 50.00 & 62.86  \\  
& - LLM-ESR & \underline{21.66} & \underline{29.58} & \underline{32.37} & \underline{35.76} & \underline{38.59} & \underline{21.66} & \underline{35.27} & \underline{42.05} & \underline{52.55} & \underline{63.77} \\
& \cellcolor{gray!10}- \textbf{GRASP} &\cellcolor{gray!10}{\textbf{23.61}}&  
\cellcolor{gray!10}{\textbf{33.62}}&  
\cellcolor{gray!10}{\textbf{37.19}}&  
\cellcolor{gray!10}{\textbf{41.01}}&  
\cellcolor{gray!10}{\textbf{44.12}}&  
\cellcolor{gray!10}{\textbf{23.61}}&  
\cellcolor{gray!10}{\textbf{40.89}}&  
\cellcolor{gray!10}{\textbf{49.56}}&  
\cellcolor{gray!10}{\textbf{61.47}}&  
\cellcolor{gray!10}{\textbf{73.62}}\\
\cmidrule{2-12} 
& SASRec & 18.84 & 25.22 &27.60	&30.58	&33.47	&18.84	&29.83	&35.62	&44.88	&56.34\\
& - RLMRec & 17.93&24.16&26.56&29.64&32.50&17.93&28.70&34.56&44.10&55.48\\  
& - LLMInit & 19.00 & 27.40 & 30.50 & 34.08 & 37.02 & 19.00 & 33.51 & 41.05 & 52.14 & 63.78 \\  
& - LLM-ESR & \underline{20.73} & \underline{29.73} & \underline{33.10} & \underline{36.99} & \underline{40.26} & \underline{20.73} & \underline{36.27} & \underline{44.49} & \underline{56.50} & \underline{69.44} \\
& \cellcolor{gray!10}- \textbf{GRASP} & \cellcolor{gray!10}{\textbf{26.56}}&  
\cellcolor{gray!10}{\textbf{36.18}}&  
\cellcolor{gray!10}{\textbf{39.33}}&  
\cellcolor{gray!10}{\textbf{42.76}}&  
\cellcolor{gray!10}{\textbf{45.61}}&  
\cellcolor{gray!10}{\textbf{26.56}}&  
\cellcolor{gray!10}{\textbf{43.09}}&  
\cellcolor{gray!10}{\textbf{50.74}}&  
\cellcolor{gray!10}{\textbf{61.33}}&  
\cellcolor{gray!10}{\textbf{72.62}}\\

\midrule[1pt]

\multirow{15}{*}{\textbf{Fashion}} 
& GRU4Rec &  25.84 & 32.71 & 34.65 & 36.66 & 38.31 & 25.84 & 37.63 & 42.37 & 48.58 & 55.07\\
& - RLMRec &29.04&34.20&35.75&37.75&39.81&29.04&37.76&41.56&47.74&55.91\\  
& - LLMInit & 33.31 & 37.48 & 38.71 & 40.29 & 42.21 & 33.31 & 40.32 & 43.31 & 48.26 & 55.89 \\  
& - LLM-ESR & \underline{37.90} & \underline{42.11} & \underline{43.42} & \underline{45.43} & \underline{47.38} & \underline{37.90} & \underline{45.03} & \underline{48.24} & \underline{54.43} & \underline{62.17} \\
& \cellcolor{gray!10}- \textbf{GRASP} & \cellcolor{gray!10}{\textbf{38.39}}&  
\cellcolor{gray!10}{\textbf{42.88}}&  
\cellcolor{gray!10}{\textbf{44.40}}&  
\cellcolor{gray!10}{\textbf{46.41}}&  
\cellcolor{gray!10}{\textbf{48.51}}&  
\cellcolor{gray!10}{\textbf{38.39}}&  
\cellcolor{gray!10}{\textbf{46.06}}&  
\cellcolor{gray!10}{\textbf{49.77}}&  
\cellcolor{gray!10}{\textbf{56.01}}&  
\cellcolor{gray!10}{\textbf{64.36}}\\
\cmidrule{2-12} 
& BERT4Rec & 28.61 & 32.00 & 33.37 & 35.58 & 37.76 & 28.61 & 34.39 & 37.74 & 44.68 & 53.23 \\
& - RLMRec & 26.95&31.92&33.40&35.41&37.36&26.95&35.33&38.95&45.16&52.91\\  
& - LLMInit & 33.99 & 37.84 & 38.92 & 40.62 & 42.43 & 33.99 & 40.48 & 43.12 & 48.42 & 55.67 \\ 
& - LLM-ESR & \textbf{37.70} & \underline{42.37} & \underline{43.75} & \underline{45.43} & \underline{47.19} & \textbf{37.70} & \underline{45.70} & \underline{49.04} & \underline{54.26} & \underline{61.26} \\
& \cellcolor{gray!10}- \textbf{GRASP} & \cellcolor{gray!10}{\underline{37.11}}&  
\cellcolor{gray!10}{\textbf{42.38}}&  
\cellcolor{gray!10}{\textbf{43.97}}&  
\cellcolor{gray!10}{\textbf{46.22}}&  
\cellcolor{gray!10}{\textbf{48.36}}&  
\cellcolor{gray!10}{\underline{37.11}}&  
\cellcolor{gray!10}{\textbf{46.09}}&  
\cellcolor{gray!10}{\textbf{50.00}}&  
\cellcolor{gray!10}{\textbf{57.01}}&  
\cellcolor{gray!10}{\textbf{65.46}}\\
\cmidrule{2-12} 
& SASRec & 39.32 & 41.93 & 42.84 & 44.13 & 45.64 & 39.32 & 43.75 & 45.95 & 49.95 & 55.98 \\
& - RLMRec & 39.94&41.96&42.72&43.92&45.32&39.94&43.40&45.26&48.98&54.55\\  
& - LLMInit & 38.91 & 42.52 & 44.04 & 46.66 & 48.27 & 38.91 & 45.68 & 49.80 & 55.32 & 61.92 \\ 
& - LLM-ESR & \underline{39.93} & \underline{43.92} & \underline{45.29} & \underline{47.15} & \underline{49.17} & \underline{39.93} & \underline{46.79} & \underline{50.13} & \underline{55.92} & \underline{64.02} \\
& \cellcolor{gray!10}- \textbf{GRASP} & \cellcolor{gray!10}{\textbf{42.16}}&  
\cellcolor{gray!10}{\textbf{46.92}}&  
\cellcolor{gray!10}{\textbf{48.50}}&  
\cellcolor{gray!10}{\textbf{50.50}}&  
\cellcolor{gray!10}{\textbf{52.57}}&  
\cellcolor{gray!10}{\textbf{42.16}}&  
\cellcolor{gray!10}{\textbf{50.30}}&  
\cellcolor{gray!10}{\textbf{54.15}}&  
\cellcolor{gray!10}{\textbf{60.31}}&  
\cellcolor{gray!10}{\textbf{68.57}}  \\

\midrule[1pt]

\multirow{15}{*}{\textbf{Industry-100K}} 
& GRU4Rec &  4.78 & 6.68 & 7.78 & 9.58 & 11.18 & 4.78 & 8.07 & 10.76 & 16.36 & 22.70 \\
& - RLMRec & 4.10&6.13&7.27&8.92&10.42&4.10&7.65&10.42&15.56&21.50\\  
& - LLMInit & 4.55 & 8.64 & 10.98 & 14.57 & 18.66 & 4.55 & 11.74 & 17.44 & 28.60 & 44.89 \\ 
& - LLM-ESR &  \underline{11.84}&\underline{18.32}&\underline{21.25}&\underline{25.13}&\underline{28.92}&\underline{11.84}&\underline{23.10}&\underline{30.25}&\underline{42.28}&\underline{57.32} \\
& - \cellcolor{gray!10}{\textbf{GRASP}} & \cellcolor{gray!10}{\textbf{13.02}} &
\cellcolor{gray!10}{\textbf{21.04}} &
\cellcolor{gray!10}{\textbf{24.54}} &
\cellcolor{gray!10}{\textbf{28.90}} &
\cellcolor{gray!10}{\textbf{32.73}} &
\cellcolor{gray!10}{\textbf{13.02}} &
\cellcolor{gray!10}{\textbf{26.96}} &
\cellcolor{gray!10}{\textbf{35.47}} &
\cellcolor{gray!10}{\textbf{48.99}} &
\cellcolor{gray!10}{\textbf{64.15}} \\
\cmidrule{2-12} 
& BERT4Rec & \underline{10.66} & \underline{13.77} & 14.98 & 16.52 & 17.94 & \underline{10.66} & 16.00 & 18.97 & 23.74 & 27.94 \\
& - RLMRec & 9.75&12.95&14.18&15.64&17.03&9.75&15.28&18.25&22.79&28.30\\  
& - LLMInit & 6.83 & 11.66 & 14.13 & 17.63 & 21.55 & 6.83 & 15.27 & 21.30 & 32.17 & 47.77 \\ 
& - LLM-ESR & 7.69&13.02&\underline{15.76}&\underline{19.67}&\underline{23.85}&7.69&\underline{17.00}&\underline{23.68}&\underline{35.85}&\underline{52.50} \\
& - \cellcolor{gray!10}{\textbf{GRASP}} &  \cellcolor{gray!10}{\textbf{14.94}}&
\cellcolor{gray!10}{\textbf{22.97}}&
\cellcolor{gray!10}{\textbf{26.44}}&
\cellcolor{gray!10}{\textbf{30.82}}&
\cellcolor{gray!10}{\textbf{34.74}}&
\cellcolor{gray!10}{\textbf{14.94}}&
\cellcolor{gray!10}{\textbf{28.86}}&
\cellcolor{gray!10}{\textbf{37.30}}&
\cellcolor{gray!10}{\textbf{50.88}}&
\cellcolor{gray!10}{\textbf{66.42}}\\
\cmidrule{2-12} 
& SASRec & 10.71 & 13.18 & 14.18 & 15.47 & 16.93 & 10.71 & 14.97 & 17.41 & 21.41 & 27.23 \\
& - RLMRec & 11.23&13.83&14.86&16.42&18.29&11.23&15.71&18.21&23.07&30.56\\  
& - LLMInit & \underline{14.99} & \underline{22.19} & \underline{25.09} & \underline{28.66} & \underline{32.06} & \underline{14.99} & \underline{27.47} & \underline{34.52} & 45.60 & 59.10 \\ 
& - LLM-ESR & 12.48&19.98&23.31&27.52&31.40&12.48&25.52&33.63&\underline{46.68}&\underline{62.04} \\
& - \cellcolor{gray!10}{\textbf{GRASP}} & \cellcolor{gray!10}{\textbf{21.81}} & \cellcolor{gray!10}{\textbf{29.05}} & \cellcolor{gray!10}{\textbf{31.69}} & \cellcolor{gray!10}{\textbf{34.88}} & \cellcolor{gray!10}{\textbf{37.96}} & \cellcolor{gray!10}{\textbf{21.81}} & \cellcolor{gray!10}{\textbf{34.27}} & \cellcolor{gray!10}{\textbf{40.70}} & \cellcolor{gray!10}{\textbf{50.60}} & \cellcolor{gray!10}{\textbf{62.84}} \\
\bottomrule[1pt]
\end{tabular}
}
\end{table*}

\begin{table*}[ht]
\centering
\caption{Comparison of model performance under different groups. (\textbf{Bold} indicates the best performance; \underline{Underline} indicates the second best performance; N = NDCG, H = Hit Rate; We report the results of the two-sided t-test with $p < 0.05$.)}
\label{tab:long_tail}
\resizebox{1\textwidth}{!}{
\renewcommand{\arraystretch}{1.2}
\footnotesize

\begin{tabular}{cc|cccc|cccc|cccc|cccc}
\toprule[1pt]

\multirow{2}{*}{\textbf{Dataset}} & \multirow{2}{*}{\textbf{Model}}  & \multicolumn{4}{c|}{\textbf{Tail User}} & \multicolumn{4}{c|}{\textbf{Tail Item}} & \multicolumn{4}{c|}{\textbf{Head User}} & \multicolumn{4}{c}{\textbf{Head Item}} \\ \cmidrule{3-18} 
 &  & \textbf{N@5} & \textbf{H@5} & \textbf{N@10} & \textbf{H@10} & \textbf{N@5} & \textbf{H@5} & \textbf{N@10} & \textbf{H@10} & \textbf{N@5} & \textbf{H@5} & \textbf{N@10} & \textbf{H@10} & \textbf{N@5} & \textbf{H@5} & \textbf{N@10} & \textbf{H@10}\\ 
 
\cmidrule{1-18}
\multirow{9}{*}{\textbf{Beauty}} 
& GRU4Rec  & 18.51 & 25.68 & 21.73 & 35.67 & 5.11 & 6.28 & 5.52 & 7.58 & 22.53 & 30.58 & 25.56 & 40.00 & 22.60 & 31.39 & 26.45 & 43.33  \\
& - LLM-ESR & \underline{27.58} & \underline{37.34} & \underline{31.26} & \underline{48.76} & \underline{6.72} & \underline{10.33} & \underline{8.61} & \underline{16.23} & \underline{30.96} & \underline{41.10} & \underline{34.35} & \underline{51.64} & \underline{33.30} & \underline{44.62} & \underline{37.35} & \underline{57.16} \\
& - \cellcolor{gray!10}\textbf{GRASP} & \cellcolor{gray!10}{\textbf{29.60}}&\cellcolor{gray!10}{\textbf{40.57}}&\cellcolor{gray!10}{\textbf{33.64}}&\cellcolor{gray!10}{\textbf{53.04}}&\cellcolor{gray!10}{\textbf{15.88}}&\cellcolor{gray!10}{\textbf{23.92}}&\cellcolor{gray!10}{\textbf{19.65}}&\cellcolor{gray!10}{\textbf{35.63}}&\cellcolor{gray!10}{\textbf{34.68}}&\cellcolor{gray!10}{\textbf{46.91}}&\cellcolor{gray!10}{\textbf{38.60}}&\cellcolor{gray!10}{\textbf{59.05}}&\cellcolor{gray!10}{\textbf{34.00}}&\cellcolor{gray!10}{\textbf{45.95}}&\cellcolor{gray!10}{\textbf{38.07}}&\cellcolor{gray!10}{\textbf{58.53}}\\
\cmidrule{2-18} 
& BERT4Rec & 18.90 & 27.34 & 22.51 & 38.54 & 0.05 & 0.12 & 0.26 & 0.76 & 23.24 & 33.40 & 27.04 & 45.11 & 24.36 & 35.18 & 28.83 & 49.01 \\
& - LLM-ESR & \underline{31.56} & \underline{41.04} & \underline{34.97} & \underline{51.59} & \underline{7.05} & \underline{9.17} & \underline{8.22} & \underline{12.83} & \underline{36.06} & \underline{46.67} & \underline{39.38} & \underline{56.94} & \underline{38.41} & \underline{49.89} & \underline{42.33} & \underline{62.03} \\
& - \cellcolor{gray!10}\textbf{GRASP} & \cellcolor{gray!10}{\textbf{36.44}}&\cellcolor{gray!10}{\textbf{48.57}}&\cellcolor{gray!10}{\textbf{40.26}}&\cellcolor{gray!10}{\textbf{60.39}}&\cellcolor{gray!10}{\textbf{14.62}}&\cellcolor{gray!10}{\textbf{22.83}}&\cellcolor{gray!10}{\textbf{18.44}}&\cellcolor{gray!10}{\textbf{34.74}}&\cellcolor{gray!10}{\textbf{40.59}}&\cellcolor{gray!10}{\textbf{54.07}}&\cellcolor{gray!10}{\textbf{44.44}}&\cellcolor{gray!10}{\textbf{65.98}}&\cellcolor{gray!10}{\textbf{42.57}}&\cellcolor{gray!10}{\textbf{55.92}}&\cellcolor{gray!10}{\textbf{46.46}}&\cellcolor{gray!10}{\textbf{67.76}}\\
\cmidrule{2-18} 
& SASRec & 26.83 & 34.52 & 29.82 & 43.78 & 5.89 & 6.90 & 6.52 & 8.89 & 31.08 & 40.63 & 34.07 & 49.88 & 32.77 & 42.46 & 36.32 & 53.46 \\
& - LLM-ESR & \underline{32.31} & \underline{43.51} & \underline{36.20} & \underline{55.56} & \underline{7.44} & \underline{12.58} & \underline{10.29} & \underline{21.47} & \underline{36.74} & \underline{48.96} & \underline{40.56} & \underline{60.79} & \underline{39.23} & \underline{52.10} & \underline{43.35} & \underline{64.85} \\
& - \cellcolor{gray!10}\textbf{GRASP} & \cellcolor{gray!10}{\textbf{38.83}}&\cellcolor{gray!10}{\textbf{49.93}}&\cellcolor{gray!10}{\textbf{42.20}}&\cellcolor{gray!10}{\textbf{60.36}}&\cellcolor{gray!10}{\textbf{23.03}}&\cellcolor{gray!10}{\textbf{31.70}}&\cellcolor{gray!10}{\textbf{26.34}}&\cellcolor{gray!10}{\textbf{41.82}}&\cellcolor{gray!10}{\textbf{41.63}}&\cellcolor{gray!10}{\textbf{54.49}}&\cellcolor{gray!10}{\textbf{45.29}}&\cellcolor{gray!10}{\textbf{65.79}}&\cellcolor{gray!10}{\textbf{43.23}}&\cellcolor{gray!10}{\textbf{55.28}}&\cellcolor{gray!10}{\textbf{46.67}}&\cellcolor{gray!10}{\textbf{65.98}} \\

\midrule[1pt]
\multirow{9}{*}{\textbf{Fashion}} 
& GRU4Rec & 22.16 & 31.00 & 24.69 & 38.79 & 0.36 & 0.70 & 0.80 & 2.12 & 50.86 & 57.11 & 52.20 & 61.28 & 48.31 & 58.96 & 50.95 & 67.08 \\
& - LLM-ESR & \underline{32.73} & \underline{38.58} & \underline{35.08} & \underline{45.82} & \underline{2.42} & \underline{3.90} & \underline{3.65} & \underline{7.76} & \underline{57.28} & \underline{60.77} & \underline{58.85} & \underline{65.60} & \underline{59.74} & \textbf{65.89} & \textbf{62.06} & \textbf{73.01} \\
& - \cellcolor{gray!10}\textbf{GRASP} & \cellcolor{gray!10}{\textbf{34.00}}&\cellcolor{gray!10}{\textbf{40.37}}&\cellcolor{gray!10}{\textbf{36.37}}&\cellcolor{gray!10}{\textbf{47.74}}&\cellcolor{gray!10}{\textbf{5.89}}&\cellcolor{gray!10}{\textbf{9.73}}&\cellcolor{gray!10}{\textbf{8.23}}&\cellcolor{gray!10}{\textbf{17.31}}&\cellcolor{gray!10}{\textbf{57.96}}&\cellcolor{gray!10}{\textbf{61.97}}&\cellcolor{gray!10}{\textbf{59.43}}&\cellcolor{gray!10}{\textbf{66.72}}&\cellcolor{gray!10}{\textbf{59.77}}&\cellcolor{gray!10}{\underline{65.71}}&\cellcolor{gray!10}{\underline{61.60}}&\cellcolor{gray!10}{\underline{71.41}}\\
\cmidrule{2-18} 
& BERT4Rec & 19.82 & 24.56 & 22.54 & 33.11 & 0.82 & 1.20 & 1.20 & 3.49 & 50.94 & 54.84 & 52.50 & 59.69 & 46.32 & 52.28 & 49.27 & 61.51 \\ 
& - LLM-ESR & \underline{32.73} & \underline{39.28} & \underline{34.67} & \underline{45.24} & \underline{1.61} & \underline{2.82} & \underline{2.57} & \underline{5.87} & \textbf{58.03} & \underline{61.71} & \underline{59.39} & \underline{65.95} & \textbf{60.52} & \underline{67.45} & \textbf{62.50} & \underline{73.52} \\
& - \cellcolor{gray!10}\textbf{GRASP} & \cellcolor{gray!10}{\textbf{33.26}}&\cellcolor{gray!10}{\textbf{40.66}}&\cellcolor{gray!10}{\textbf{35.97}}&\cellcolor{gray!10}{\textbf{48.98}}&\cellcolor{gray!10}{\textbf{3.30}}&\cellcolor{gray!10}{\textbf{5.97}}&\cellcolor{gray!10}{\textbf{5.64}}&\cellcolor{gray!10}{\textbf{13.31}}&\cellcolor{gray!10}{\underline{57.86}}&\cellcolor{gray!10}{\textbf{62.13}}&\cellcolor{gray!10}{\textbf{59.57}}&\cellcolor{gray!10}{\textbf{67.43}}&\cellcolor{gray!10}{\underline{60.16}}&\cellcolor{gray!10}{\textbf{67.53}}&\cellcolor{gray!10}{\underline{62.38}}&\cellcolor{gray!10}{\textbf{74.40}} \\
\cmidrule{2-18} 
& SASRec & 32.35 & 35.60 & 33.82 & 40.18 & 1.68 & 2.39 & 2.13 & 3.78 & 56.45 & 59.38 & 57.49 & 62.62 & 59.22 & 63.29 & 60.85 & 68.32 \\
& - LLM-ESR & \underline{35.02} & \underline{40.55} & \underline{37.31} & \underline{47.67} & \underline{3.28} & \underline{5.33} & \underline{4.96} & \underline{10.58} & \underline{58.61} & \underline{62.57} & \underline{59.90} & \underline{66.61} & \underline{62.01} & \underline{67.97} & \underline{63.94} & \textbf{73.97}\\
& - \cellcolor{gray!10}\textbf{GRASP} & \cellcolor{gray!10}{\textbf{39.53}}&\cellcolor{gray!10}{\textbf{46.25}}&\cellcolor{gray!10}{\textbf{41.85}}&\cellcolor{gray!10}{\textbf{53.44}}&\cellcolor{gray!10}{\textbf{12.28}}&\cellcolor{gray!10}{\textbf{17.79}}&\cellcolor{gray!10}{\textbf{15.07}}&\cellcolor{gray!10}{\textbf{26.46}}&\cellcolor{gray!10}{\textbf{60.14}}&\cellcolor{gray!10}{\textbf{64.39}}&\cellcolor{gray!10}{\textbf{61.77}}&\cellcolor{gray!10}{\textbf{69.24}}&\cellcolor{gray!10}{\textbf{62.92}}&\cellcolor{gray!10}{\textbf{68.63}}&\cellcolor{gray!10}{\textbf{64.59}}&\cellcolor{gray!10}{\underline{73.78}}\\

\midrule[1pt]
\multirow{9}{*}{\textbf{Industry-100K}} 
& GRU4Rec & 7.89&10.96&9.65&16.41&0.48&0.82&0.71&1.53&7.69&10.66&9.37&15.91&15.86&21.85&19.25&32.38\\
& - LLM-ESR &  \underline{20.78}&\underline{29.68}&\underline{24.67}&\underline{41.74}&\underline{20.97}&\underline{29.97}&\underline{24.81}&\underline{41.87}&\underline{23.21}&\underline{32.67}&\underline{27.04}&\underline{44.53}&\underline{21.55}&\underline{30.57}&\underline{25.47}&\underline{42.73}\\
& - \cellcolor{gray!10}{\textbf{GRASP}} &
\cellcolor{gray!10}{\textbf{24.09}} &
\cellcolor{gray!10}{\textbf{34.94}} &
\cellcolor{gray!10}{\textbf{28.46}} &
\cellcolor{gray!10}{\textbf{48.47}} &
\cellcolor{gray!10}{\textbf{24.30}} &
\cellcolor{gray!10}{\textbf{35.35}} &
\cellcolor{gray!10}{\textbf{28.66}} &
\cellcolor{gray!10}{\textbf{48.87}} &
\cellcolor{gray!10}{\textbf{26.41}} &
\cellcolor{gray!10}{\textbf{37.70}} &
\cellcolor{gray!10}{\textbf{30.75}} &
\cellcolor{gray!10}{\textbf{51.14}} &
\cellcolor{gray!10}{\textbf{24.80}} &
\cellcolor{gray!10}{\textbf{35.61}} &
\cellcolor{gray!10}{\textbf{29.17}} &
\cellcolor{gray!10}{\textbf{49.12}}\\
\cmidrule{2-18} 
& BERT4Rec & 14.16 & 18.17 & 15.67 & 22.86 & 5.49 & 5.64 & 5.50 & 5.64 & \underline{17.96} & 21.60 & 19.30 & 25.76 & \underline{25.10} & \underline{33.17} & \underline{28.18} & \underline{42.73}\\ 
& - LLM-ESR & \underline{15.66}&\underline{23.63}&\underline{19.58}&\underline{35.85}&\underline{14.15}&\underline{21.81}&\underline{17.95}&\underline{33.66}&16.18&\underline{23.90}&\underline{20.02}&\underline{35.86}&17.50&25.71&21.52&38.23\\
& - \cellcolor{gray!10}{\textbf{GRASP}} & \cellcolor{gray!10}{\textbf{26.64}}&
\cellcolor{gray!10}{\textbf{37.54}}&
\cellcolor{gray!10}{\textbf{31.02}}&
\cellcolor{gray!10}{\textbf{51.09}}&
\cellcolor{gray!10}{\textbf{26.09}}&
\cellcolor{gray!10}{\textbf{37.07}}&
\cellcolor{gray!10}{\textbf{30.38}}&
\cellcolor{gray!10}{\textbf{50.34}}&
\cellcolor{gray!10}{\textbf{25.58}}&
\cellcolor{gray!10}{\textbf{36.30}}&
\cellcolor{gray!10}{\textbf{30.00}}&
\cellcolor{gray!10}{\textbf{49.99}}&
\cellcolor{gray!10}{\textbf{26.81}}&
\cellcolor{gray!10}{\textbf{37.56}}&
\cellcolor{gray!10}{\textbf{31.30}}&
\cellcolor{gray!10}{\textbf{51.47}} \\
\cmidrule{2-18} 
& SASRec & 13.21&16.56&14.63&20.96&5.26&5.54&5.38&5.92&17.17&20.26&18.36&23.93&23.45&30.02&26.17&38.48 \\
& - LLM-ESR & \underline{23.35}&\underline{33.65}&\underline{27.53}&\underline{46.62}&\underline{22.96}&\underline{33.40}&\underline{27.22}&\underline{46.62}&\underline{23.17}&\underline{33.55}&\underline{27.49}&\underline{46.94}&\underline{23.70}&\underline{33.88}&\underline{27.85}&\underline{46.75}\\
& - \cellcolor{gray!10}{\textbf{GRASP}} &
\cellcolor{gray!10}{\textbf{31.64}} &
\cellcolor{gray!10}{\textbf{40.26}} &
\cellcolor{gray!10}{\textbf{34.76}} &
\cellcolor{gray!10}{\textbf{49.94}} &
\cellcolor{gray!10}{\textbf{32.39}} &
\cellcolor{gray!10}{\textbf{41.53}} &
\cellcolor{gray!10}{\textbf{35.45}} &
\cellcolor{gray!10}{\textbf{51.02}} &
\cellcolor{gray!10}{\textbf{31.87}} &
\cellcolor{gray!10}{\textbf{42.51}} &
\cellcolor{gray!10}{\textbf{35.38}} &
\cellcolor{gray!10}{\textbf{53.38}} &
\cellcolor{gray!10}{\textbf{30.92}} &
\cellcolor{gray!10}{\textbf{39.79}} &
\cellcolor{gray!10}{\textbf{34.26}} &
\cellcolor{gray!10}{\textbf{50.15}} \\ 
\bottomrule[1pt]
\end{tabular}
}
\end{table*}

\begin{figure*}[t]
\vspace{-8pt}
\centerline{\includegraphics[width=\linewidth]{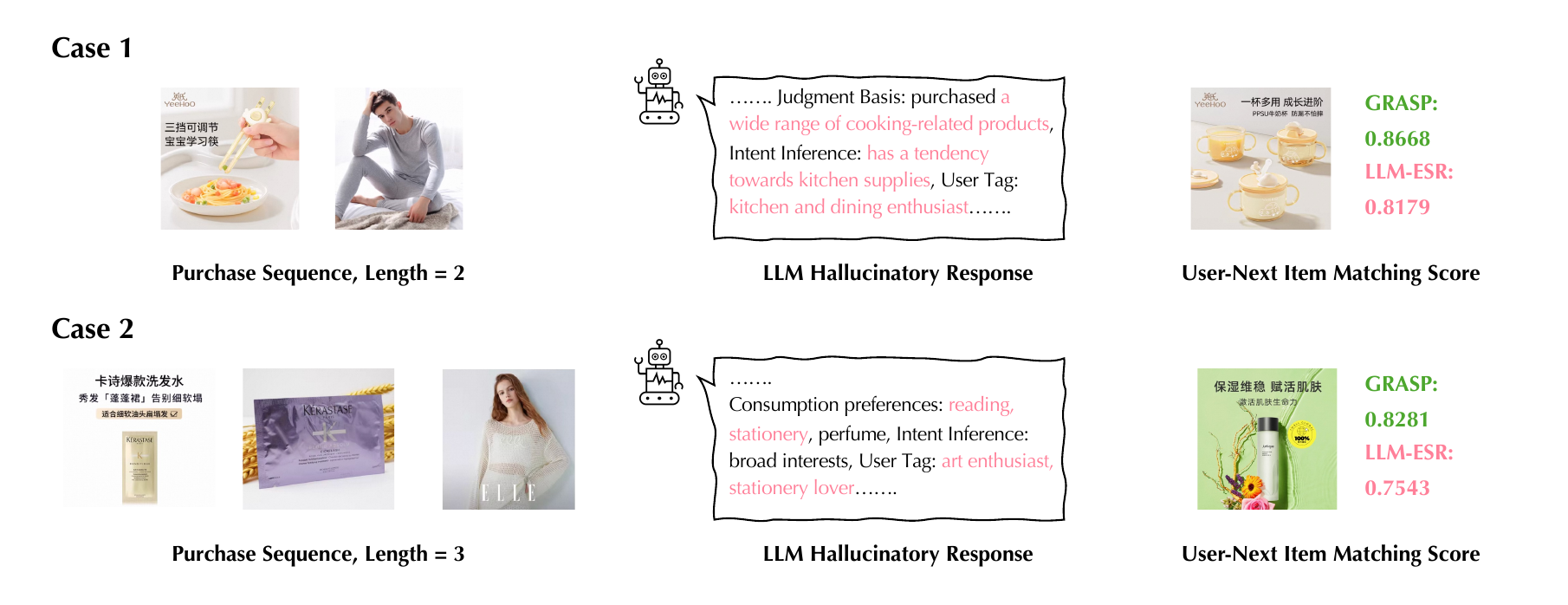}}
\caption{Examples of LLM hallucinations in Industry-100K and comparison of recommendation results from different models.}
\label{fig:visual_hall}
\end{figure*}

\begin{table*}[t]
\centering
\caption{Ablation study on Beauty dataset on the basis of SASRec. (\textbf{Bold} indicates the best performance; \underline{Underline} indicates the second best performance; N = NDCG, H = Hit Rate)}
\label{tab:ablation}
\resizebox{1\textwidth}{!}{
\renewcommand{\arraystretch}{1.2} 
\footnotesize
\begin{tabular}{cc|ccccc|ccccc}
\toprule[1pt]
\textbf{Module} & \textbf{Setting} & \textbf{N@1} & \textbf{N@3} & \textbf{N@5} & \textbf{N@10} & \textbf{N@20} & \textbf{H@1} & \textbf{H@3} & \textbf{H@5} & \textbf{H@10} & \textbf{H@20} \\ 
\midrule

\multirow{4}{*}{HAE} 
& - \textit{w/o} Attention & 18.24 & 26.63 & 29.83 & 33.41 & 36.70 & 18.24 & 32.71 & 40.48 & 51.59 & 64.63 \\ 
& - \textit{w/o} $\mathbf{i}_{similar}^{\text{HAE}}$ & 18.74 & 27.16 & 30.48 & 34.35 & 37.75 & 18.74 & 33.29 & 41.36 & 53.35 & 66.83 \\
& - \textit{w/o} $\mathbf{i}_{global}^{\text{HAE}}$ & 20.00 & 29.29 & 32.72 & 36.62 & 39.87 & 20.00 & 36.02 & 44.36 & 56.44 & 69.31 \\
& - $\text{Softmax}$ & 14.59 & 22.60 & 25.92 & 30.00 & 33.63 & 14.59 & 28.46 & 36.54 & 49.17 & 63.55 \\

\midrule

\multicolumn{2}{c|}{\cellcolor{gray!10}\textbf{GRASP}} &  \cellcolor{gray!10}{\textbf{26.56}}&  
\cellcolor{gray!10}{\textbf{36.18}}&  
\cellcolor{gray!10}{\textbf{39.33}}&  
\cellcolor{gray!10}{\textbf{42.76}}&  
\cellcolor{gray!10}{\textbf{45.61}}&  
\cellcolor{gray!10}{\textbf{26.56}}&  
\cellcolor{gray!10}{\textbf{43.09}}&  
\cellcolor{gray!10}{\textbf{50.74}}&  
\cellcolor{gray!10}{\textbf{61.33}}&  
\cellcolor{gray!10}{\textbf{72.62}}\\
					
\bottomrule[1pt]
\end{tabular}
\vspace{-3pt}

}
\end{table*}

\begin{figure*}[t]
\centerline{\includegraphics[width=0.9\linewidth]{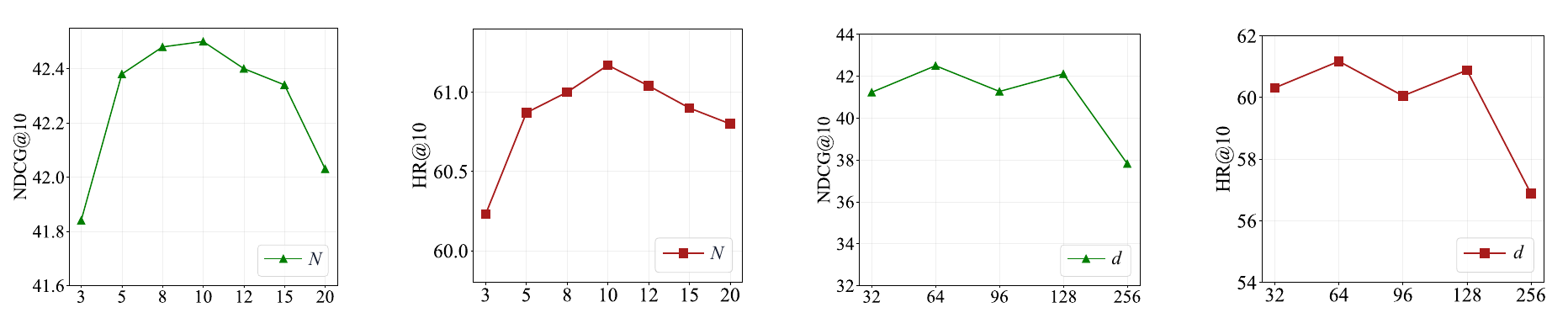}}
\caption{Analysis of hyper-parameters on
Beauty dataset of GRASP based on SASRec. \textbf{Left}: $N$ is the size of the candidate pool for similar retrieval. \textbf{Right}: $d$ is the hidden dimension for SRS.
}
\label{fig:hyper-parameter}
\end{figure*}


\subsection{Performance Analysis}

\textbf{Overall performance}. As shown in Table~\ref{tab:main_exp}, our method consistently outperforms other state-of-the-art models on three sequential recommendation benchmarks. On the Beauty dataset, our approach achieves an average improvement of 4.56\% over the previous best-performing model, LLM-ESR. Similarly, on the Fashion dataset, GRASP surpasses LLM-ESR by 1.81\%, and on the large-scale internal dataset Industry-100K, it achieves a substantial gain of 6.68\%. In addition to outperforming previous LLM4Rec-related works, GRASP demonstrates strong compatibility when integrated with various sequential recommendation architectures, including GRU4Rec, BERT4Rec, and SASRec, consistently delivering performance boosts. This further illustrates the flexibility and transferability of the proposed framework, making it a promising solution for a wide range of sequential recommendation tasks.

\textbf{Performance under different groups}. In addition to evaluating overall performance, we also analyze GRASP's effectiveness in different scenarios. As highlighted in Section~\ref{sec:intro}, existing LLM-based sequential recommendation models often suffer from hallucination issues, especially in the \textit{tail} part. To thoroughly compare models' performances, we delineate tail users and tail items based on the Pareto Principle, where users and items with interaction frequencies in the top 20\% are classified as \textit{head}, while the remaining are defined as \textit{tail}~\cite{liu2024llm, box1986analysis}. As shown in Table~\ref{tab:long_tail}, we compare GRASP against the SOTA LLM-ESR and three basic sequential recommendation models. In \textit{tail} scenarios—where hallucination risks are highest due to sparse interactions—GRASP consistently outperforms LLM-ESR and all baseline models across all datasets. On the Fashion dataset, GRASP surpasses LLM-ESR by an average of 5.00\%. The improvement is even more pronounced on the Beauty dataset, where GRASP demonstrates the most substantial enhancement over LLM-ESR in all cases, achieving a remarkable 9.99\% increase. Notably, on the real-world Industry-100K dataset, GRASP also demonstrates a significant improvement, outperforming LLM-ESR or SRS baselines by an impressive 8.42\%. Besides, GRASP maintains strong performance in \textit{head} scenarios, where interaction data is abundant and hallucinations are less prevalent. Specifically, GRASP improves over SOTA by 0.57\% on the Fashion dataset, 4.30\% on the Beauty dataset, and 6.41\% on the Industry-100K dataset. This ensures that improvements in long-tail scenarios do not come at the expense of head performance. This balanced performance underscores GRASP's ability to mitigate hallucination effects in data-scarce scenarios while preserving high recommendation accuracy for well-represented users and items. 

Furthermore, we select two examples from Industry-100K where LLMs exhibit hallucinated descriptions, along with the corresponding ground truth next-item and the user interaction score, obtained by applying sigmoid function to the dot product of their embeddings. The hallucinated descriptions may introduce noise for user preference demonstration, and then adding interference in learning user interests. In Fig.~\ref{fig:visual_hall}, it can be observed that GRASP align better with user expectations than LLM-ESR, further demonstrating GRASP's robustness against hallucination issues.

\subsection{Ablation Study}
\textbf{Ablation on each component}. We demonstrate the effect of each component in GRASP. In the absence of \textit{generation augmented retrieval}, the model would degenerate into the SRS baselines. Specifically, we remove each part in \textit{holistic attention enhancement}~(\ie attention, $\mathbf{i}_{similar}^{\text{HAE}}$, $\mathbf{i}_{global}^{\text{HAE}}$, replace $\text{Sigmoid}$ with $\text{Softmax}$), respectively. As shown in Table~\ref{tab:ablation}, it is evident that removing any part in \textit{holistic attention enhancement} has a relatively significant degradation on performance, highlighting the efficacy of our fine-grained user-item integration together with global content interaction. Furthermore, replacing the traditional softmax function with sigmoid in the attention mechanism proves to be a critical enhancement, as it preserves the independence of items within a sequence while enabling fine-grained and contextually precise feature enhancement.

\textbf{Impacts of hyper-parameters}. As shown in Fig.~\ref{fig:hyper-parameter}, the model’s performance is related to the choice of $N$ and $d$, which represent the size of the candidate pool used for retrieving similar users/items and the hidden embedding dimension for SRS. For the candidate pool size $N$, we observe that values that are too small fail to capture sufficient similar patterns, while excessively large values introduce noise and irrelevant information. Regarding embedding dimension $d$, the results indicate that insufficient dimensionality cannot adequately represent complex user-item relationships, while excessive dimensionality leads to diminished returns and potential overfitting. Based on the results, the optimal values are $N = 10$ and $d = 64$, ensuring relatively superior performance.

%% file: 5Conclusion.tex
\section{Conclusion and Future Work}
In this paper, we propose \textbf{GRASP}, a novel framework that empowers the sequential recommendation model by synergizing generation augmented retrieval with holistic attention enhancement. Unlike existing methods that suffer from sparse supervision signals or directly utilizing potentially unreliable semantic features as supervision, GRASP leverages descriptive synthesis and similarity retrieval to provide robust auxiliary information, while multi-level attention mechanisms dynamically enrich user-item representations with contextual signals from similar users/items—effectively circumventing noisy guidance. Comprehensive experiments on three datasets demonstrate GRASP's consistent superiority over state-of-the-art baselines, offering a flexible solution adaptable to diverse sequential recommendation scenarios.

Furthermore, our GRASP utilizes LLM's world knowledge as a front feature augmentation module, which is easy to deploy in real-world applications. In the meanwhile, some LLM based recommendation works~\cite{zheng2024adapting,wang2025act} inherently use the pre-trained weights of LLM in the SRS backbone. This technique is orthogonal to ours, and we will explore how to combine them together for more impressive model performance in the future. 


%% file: main.bib
@article{xu2024prompting,
  title={Prompting large language models for recommender systems: A comprehensive framework and empirical analysis},
  author={Xu, Lanling and Zhang, Junjie and Li, Bingqian and Wang, Jinpeng and Cai, Mingchen and Zhao, Wayne Xin and Wen, Ji-Rong},
  journal={arXiv preprint arXiv:2401.04997},
  year={2024}
}

@inproceedings{bao2023tallrec,
  title={Tallrec: An effective and efficient tuning framework to align large language model with recommendation},
  author={Bao, Keqin and Zhang, Jizhi and Zhang, Yang and Wang, Wenjie and Feng, Fuli and He, Xiangnan},
  booktitle={Proceedings of the 17th ACM Conference on Recommender Systems},
  pages={1007--1014},
  year={2023}
}

@article{liu2024llm,
  title={Llm-esr: Large language models enhancement for long-tailed sequential recommendation},
  author={Liu, Qidong and Wu, Xian and Wang, Yejing and Zhang, Zijian and Tian, Feng and Zheng, Yefeng and Zhao, Xiangyu},
  journal={Advances in Neural Information Processing Systems},
  volume={37},
  pages={26701--26727},
  year={2024}
}

@inproceedings{yang2024sequential,
  title={Sequential recommendation with latent relations based on large language model},
  author={Yang, Shenghao and Ma, Weizhi and Sun, Peijie and Ai, Qingyao and Liu, Yiqun and Cai, Mingchen and Zhang, Min},
  booktitle={Proceedings of the 47th International ACM SIGIR Conference on Research and Development in Information Retrieval},
  pages={335--344},
  year={2024}
}

@inproceedings{hidasi2015session,
  title={Session-based recommendations with recurrent neural networks},
  author={Hidasi, Bal{\'a}zs and Karatzoglou, Alexandros and Baltrunas, Linas and Tikk, Domonkos},
  booktitle={The International Conference on Learning Representations},
  year={2016}
}

@inproceedings{sun2019bert4rec,
  title={BERT4Rec: Sequential recommendation with bidirectional encoder representations from transformer},
  author={Sun, Fei and Liu, Jun and Wu, Jian and Pei, Changhua and Lin, Xiao and Ou, Wenwu and Jiang, Peng},
  booktitle={Proceedings of the 28th ACM international conference on information and knowledge management},
  pages={1441--1450},
  year={2019}
}

@inproceedings{kang2018self,
  title={Self-attentive sequential recommendation},
  author={Kang, Wang-Cheng and McAuley, Julian},
  booktitle={2018 IEEE international conference on data mining (ICDM)},
  pages={197--206},
  year={2018},
  organization={IEEE}
}

@inproceedings{ren2024representation,
  title={Representation learning with large language models for recommendation},
  author={Ren, Xubin and Wei, Wei and Xia, Lianghao and Su, Lixin and Cheng, Suqi and Wang, Junfeng and Yin, Dawei and Huang, Chao},
  booktitle={Proceedings of the ACM on Web Conference 2024},
  pages={3464–3475},
  year={2024}
}

@inproceedings{harte2023leveraging,
  title={Leveraging large language models for sequential recommendation},
  author={Harte, Jesse and Zorgdrager, Wouter and Louridas, Panos and Katsifodimos, Asterios and Jannach, Dietmar and Fragkoulis, Marios},
  booktitle={Proceedings of the 17th ACM Conference on Recommender Systems},
  pages={1096--1102},
  year={2023}
}

@inproceedings{hu2024enhancing,
  title={Enhancing Sequential Recommendation via LLM-based Semantic Embedding Learning},
  author={Hu, Jun and Xia, Wenwen and Zhang, Xiaolu and Fu, Chilin and Wu, Weichang and Huan, Zhaoxin and Li, Ang and Tang, Zuoli and Zhou, Jun},
  booktitle={Companion Proceedings of the ACM on Web Conference 2024},
  pages={103--111},
  year={2024}
}

@article{yang2024qwen2,
  title={Qwen2. 5 technical report},
  author={Yang, An and Yang, Baosong and Zhang, Beichen and Hui, Binyuan and Zheng, Bo and Yu, Bowen and Li, Chengyuan and Liu, Dayiheng and Huang, Fei and Wei, Haoran and others},
  journal={arXiv preprint arXiv:2412.15115},
  year={2024}
}

@article{behnamghader2024llm2vec,
  title={Llm2vec: Large language models are secretly powerful text encoders},
  author={BehnamGhader, Parishad and Adlakha, Vaibhav and Mosbach, Marius and Bahdanau, Dzmitry and Chapados, Nicolas and Reddy, Siva},
  journal={arXiv preprint arXiv:2404.05961},
  year={2024}
}

@inproceedings{mcauley2015image,
  title={Image-based recommendations on styles and substitutes},
  author={McAuley, Julian and Targett, Christopher and Shi, Qinfeng and Van Den Hengel, Anton},
  booktitle={Proceedings of the 38th international ACM SIGIR conference on research and development in information retrieval},
  pages={43--52},
  year={2015}
}

@article{box1986analysis,
  title={An analysis for unreplicated fractional factorials},
  author={Box, George EP and Meyer, R Daniel},
  journal={Technometrics},
  volume={28},
  number={1},
  pages={11--18},
  year={1986},
  publisher={Taylor \& Francis}
}

@inproceedings{xie2022contrastive,
  title={Contrastive learning for sequential recommendation},
  author={Xie, Xu and Sun, Fei and Liu, Zhaoyang and Wu, Shiwen and Gao, Jinyang and Zhang, Jiandong and Ding, Bolin and Cui, Bin},
  booktitle={2022 IEEE 38th international conference on data engineering (ICDE)},
  pages={1259--1273},
  year={2022},
  organization={IEEE}
}

@inproceedings{chen2018sequential,
  title={Sequential recommendation with user memory networks},
  author={Chen, Xu and Xu, Hongteng and Zhang, Yongfeng and Tang, Jiaxi and Cao, Yixin and Qin, Zheng and Zha, Hongyuan},
  booktitle={Proceedings of the eleventh ACM international conference on web search and data mining},
  pages={108--116},
  year={2018}
}

@article{boka2024survey,
  title={A survey of sequential recommendation systems: Techniques, evaluation, and future directions},
  author={Boka, Tesfaye Fenta and Niu, Zhendong and Neupane, Rama Bastola},
  journal={Information Systems},
  pages={102427},
  year={2024},
  publisher={Elsevier}
}

@inproceedings{yang2023debiased,
  title={Debiased contrastive learning for sequential recommendation},
  author={Yang, Yuhao and Huang, Chao and Xia, Lianghao and Huang, Chunzhen and Luo, Da and Lin, Kangyi},
  booktitle={Proceedings of the ACM web conference 2023},
  pages={1063--1073},
  year={2023}
}

@article{nasir2023survey,
  title={A survey and taxonomy of sequential recommender systems for e-commerce product recommendation},
  author={Nasir, Mahreen and Ezeife, Christie I},
  journal={SN Computer Science},
  volume={4},
  number={6},
  pages={708},
  year={2023},
  publisher={Springer}
}

@article{zheng2024large,
  title={A large language model enhanced sequential recommender for joint video and comment recommendation},
  author={Zheng, Bowen and Lin, Zihan and Liu, Enze and Yang, Chen and Bai, Enyang and Ling, Cheng and Zhao, Wayne Xin and Wen, Ji-Rong},
  journal={arXiv preprint arXiv:2403.13574},
  year={2024}
}

@inproceedings{gao2022kuairand,
  title={Kuairand: An unbiased sequential recommendation dataset with randomly exposed videos},
  author={Gao, Chongming and Li, Shijun and Zhang, Yuan and Chen, Jiawei and Li, Biao and Lei, Wenqiang and Jiang, Peng and He, Xiangnan},
  booktitle={Proceedings of the 31st ACM International Conference on Information \& Knowledge Management},
  pages={3953--3957},
  year={2022}
}

@inproceedings{li2024calrec,
  title={Calrec: Contrastive alignment of generative llms for sequential recommendation},
  author={Li, Yaoyiran and Zhai, Xiang and Alzantot, Moustafa and Yu, Keyi and Vuli{\'c}, Ivan and Korhonen, Anna and Hammad, Mohamed},
  booktitle={Proceedings of the 18th ACM Conference on Recommender Systems},
  pages={422--432},
  year={2024}
}

@article{boz2024improving,
  title={Improving sequential recommendations with llms},
  author={Boz, Artun and Zorgdrager, Wouter and Kotti, Zoe and Harte, Jesse and Louridas, Panos and Karakoidas, Vassilios and Jannach, Dietmar and Fragkoulis, Marios},
  journal={ACM Transactions on Recommender Systems},
  year={2024},
  publisher={ACM New York, NY}
}

@inproceedings{acharya2023llm,
  title={Llm based generation of item-description for recommendation system},
  author={Acharya, Arkadeep and Singh, Brijraj and Onoe, Naoyuki},
  booktitle={Proceedings of the 17th ACM conference on recommender systems},
  pages={1204--1207},
  year={2023}
}

@inproceedings{kim2024large,
  title={Large language models meet collaborative filtering: An efficient all-round llm-based recommender system},
  author={Kim, Sein and Kang, Hongseok and Choi, Seungyoon and Kim, Donghyun and Yang, Minchul and Park, Chanyoung},
  booktitle={Proceedings of the 30th ACM SIGKDD Conference on Knowledge Discovery and Data Mining},
  pages={1395--1406},
  year={2024}
}

@article{fan2024recommender,
  title={Recommender systems in the era of large language models (llms)},
  author={Fan, Wenqi},
  journal={IEEE Transactions on Knowledge and Data Engineering},
  pages={1--20},
  year={2024},
  publisher={IEEE Computer Society}
}

@inproceedings{liu2025llmemb,
  title={LLMEmb: Large Language Model Can Be a Good Embedding Generator for Sequential Recommendation},
  author={Liu, Qidong and Wu, Xian and Wang, Wanyu and Wang, Yejing and Zhu, Yuanshao and Zhao, Xiangyu and Tian, Feng and Zheng, Yefeng},
  booktitle={Proceedings of the AAAI Conference on Artificial Intelligence},
  pages={12183--12191},
  year={2025}
}

@inproceedings{jia2025learn,
  title={LEARN: Knowledge Adaptation from Large Language Model to Recommendation for Practical Industrial Application},
  author={Jia, Jian and Wang, Yipei and Li, Yan and Chen, Honggang and Bai, Xuehan and Liu, Zhaocheng and Liang, Jian and Chen, Quan and Li, Han and Jiang, Peng and others},
  booktitle={Proceedings of the AAAI Conference on Artificial Intelligence},
  pages={11861--11869},
  year={2025}
}

@article{quadrana2018sequence,
  title={Sequence-aware recommender systems},
  author={Quadrana, Massimo and Cremonesi, Paolo and Jannach, Dietmar},
  journal={ACM computing surveys (CSUR)},
  volume={51},
  number={4},
  pages={1--36},
  year={2018},
  publisher={ACM New York, NY, USA}
}

@article{mishra2015web,
  title={A web recommendation system considering sequential information},
  author={Mishra, Rajhans and Kumar, Pradeep and Bhasker, Bharat},
  journal={Decision Support Systems},
  volume={75},
  pages={1--10},
  year={2015},
  publisher={Elsevier}
}

@article{wang2019sequential,
  title={Sequential recommender systems: challenges, progress and prospects},
  author={Wang, Shoujin and Hu, Liang and Wang, Yan and Cao, Longbing and Sheng, Quan Z and Orgun, Mehmet},
  journal={arXiv preprint arXiv:2001.04830},
  year={2019}
}

@inproceedings{ying2018sequential,
  title={Sequential recommender system based on hierarchical attention network},
  author={Ying, Haochao and Zhuang, Fuzhen and Zhang, Fuzheng and Liu, Yanchi and Xu, Guandong and Xie, Xing and Xiong, Hui and Wu, Jian},
  booktitle={IJCAI international joint conference on artificial intelligence},
  year={2018}
}

@article{chen2025enhancing,
  title={Enhancing ID-based Recommendation with Large Language Models},
  author={Chen, Lei and Gao, Chen and Du, Xiaoyi and Luo, Hengliang and Jin, Depeng and Li, Yong and Wang, Meng},
  journal={ACM Transactions on Information Systems},
  volume={43},
  number={5},
  pages={1--30},
  year={2025},
  publisher={ACM New York, NY}
}

@inproceedings{yuan2023go,
  title={Where to go next for recommender systems? id-vs. modality-based recommender models revisited},
  author={Yuan, Zheng and Yuan, Fajie and Song, Yu and Li, Youhua and Fu, Junchen and Yang, Fei and Pan, Yunzhu and Ni, Yongxin},
  booktitle={Proceedings of the 46th International ACM SIGIR Conference on Research and Development in Information Retrieval},
  pages={2639--2649},
  year={2023}
}

@article{pan2024survey,
  title={A Survey on Sequential Recommendation},
  author={Pan, Liwei and Pan, Weike and Wei, Meiyan and Yin, Hongzhi and Ming, Zhong},
  journal={arXiv preprint arXiv:2412.12770},
  year={2024}
}

@article{li2025id,
  title={From ID-based to ID-free: Rethinking ID Effectiveness in Multimodal Collaborative Filtering Recommendation},
  author={Li, Guohao and Jing, Li and Wu, Jia and Li, Xuefei and Zhu, Kai and He, Yue},
  journal={arXiv preprint arXiv:2507.05715},
  year={2025}
}

@article{huang2025survey,
  title={A survey on hallucination in large language models: Principles, taxonomy, challenges, and open questions},
  author={Huang, Lei and Yu, Weijiang and Ma, Weitao and Zhong, Weihong and Feng, Zhangyin and Wang, Haotian and Chen, Qianglong and Peng, Weihua and Feng, Xiaocheng and Qin, Bing and others},
  journal={ACM Transactions on Information Systems},
  volume={43},
  number={2},
  pages={1--55},
  year={2025},
  publisher={ACM New York, NY}
}

@article{rawte2023survey,
  title={A survey of hallucination in large foundation models},
  author={Rawte, Vipula and Sheth, Amit and Das, Amitava},
  journal={arXiv preprint arXiv:2309.05922},
  year={2023}
}

@inproceedings{chang2021sequential,
  title={Sequential recommendation with graph neural networks},
  author={Chang, Jianxin and Gao, Chen and Zheng, Yu and Hui, Yiqun and Niu, Yanan and Song, Yang and Jin, Depeng and Li, Yong},
  booktitle={Proceedings of the 44th international ACM SIGIR conference on research and development in information retrieval},
  pages={378--387},
  year={2021}
}

@inproceedings{liu2016context,
  title={Context-aware sequential recommendation},
  author={Liu, Qiang and Wu, Shu and Wang, Diyi and Li, Zhaokang and Wang, Liang},
  booktitle={2016 IEEE 16th International Conference on Data Mining (ICDM)},
  pages={1053--1058},
  year={2016},
  organization={IEEE}
}

@inproceedings{tang2018personalized,
  title={Personalized top-n sequential recommendation via convolutional sequence embedding},
  author={Tang, Jiaxi and Wang, Ke},
  booktitle={Proceedings of the eleventh ACM international conference on web search and data mining},
  pages={565--573},
  year={2018}
}

@inproceedings{xu2019recurrent,
  title={Recurrent convolutional neural network for sequential recommendation},
  author={Xu, Chengfeng and Zhao, Pengpeng and Liu, Yanchi and Xu, Jiajie and S. Sheng, Victor S Sheng and Cui, Zhiming and Zhou, Xiaofang and Xiong, Hui},
  booktitle={The world wide web conference},
  pages={3398--3404},
  year={2019}
}

@inproceedings{jiang2024reformulating,
  title={Reformulating sequential recommendation: Learning dynamic user interest with content-enriched language modeling},
  author={Jiang, Junzhe and Qu, Shang and Cheng, Mingyue and Liu, Qi and Liu, Zhiding and Zhang, Hao and Zhang, Rujiao and Zhang, Kai and Li, Rui and Li, Jiatong and others},
  booktitle={International Conference on Database Systems for Advanced Applications},
  pages={353--362},
  year={2024},
  organization={Springer}
}

@article{wang2024towards,
  title={Towards next-generation llm-based recommender systems: A survey and beyond},
  author={Wang, Qi and Li, Jindong and Wang, Shiqi and Xing, Qianli and Niu, Runliang and Kong, He and Li, Rui and Long, Guodong and Chang, Yi and Zhang, Chengqi},
  journal={arXiv preprint arXiv:2410.19744},
  year={2024}
}

@article{gao2024llm,
  title={Llm-enhanced reranking in recommender systems},
  author={Gao, Jingtong and Chen, Bo and Zhao, Xiangyu and Liu, Weiwen and Li, Xiangyang and Wang, Yichao and Zhang, Zijian and Wang, Wanyu and Ye, Yuyang and Lin, Shanru and others},
  journal={arXiv preprint arXiv:2406.12433},
  year={2024}
}

@inproceedings{wang2024llm4msr,
  title={Llm4msr: An llm-enhanced paradigm for multi-scenario recommendation},
  author={Wang, Yuhao and Wang, Yichao and Fu, Zichuan and Li, Xiangyang and Wang, Wanyu and Ye, Yuyang and Zhao, Xiangyu and Guo, Huifeng and Tang, Ruiming},
  booktitle={Proceedings of the 33rd ACM International Conference on Information and Knowledge Management},
  pages={2472--2481},
  year={2024}
}

@inproceedings{zhang2025does,
  title={How does Search Affect Personalized Recommendations and User Behavior? Evidence from LLM-based Synthetic Data},
  author={Zhang, Haoran and Kang, Xin and Guo, Junpeng},
  booktitle={Companion Proceedings of the ACM on Web Conference 2025},
  pages={2434--2443},
  year={2025}
}

@article{feng2025iranker,
  title={IRanker: Towards Ranking Foundation Model},
  author={Feng, Tao and Hua, Zhigang and Lei, Zijie and Xie, Yan and Yang, Shuang and Long, Bo and You, Jiaxuan},
  journal={arXiv preprint arXiv:2506.21638},
  year={2025}
}

@article{HLLM,
title={HLLM: Enhancing Sequential Recommendations via Hierarchical Large Language Models for Item and User Modeling},
author={Junyi Chen and Lu Chi and Bingyue Peng and Zehuan Yuan},
journal={arXiv preprint arXiv:2409.12740},
year={2024}
}

@inproceedings{zheng2024adapting,
  title={Adapting large language models by integrating collaborative semantics for recommendation},
  author={Zheng, Bowen and Hou, Yupeng and Lu, Hongyu and Chen, Yu and Zhao, Wayne Xin and Chen, Ming and Wen, Ji-Rong},
  booktitle={2024 IEEE 40th International Conference on Data Engineering (ICDE)},
  pages={1435--1448},
  year={2024},
  organization={IEEE}
}

@article{wang2025act,
  title={Act-With-Think: Chunk Auto-Regressive Modeling for Generative Recommendation},
  author={Wang, Yifan and Gan, Weinan and Xiao, Longtao and Zhu, Jieming and Chang, Heng and Wang, Haozhao and Zhang, Rui and Dong, Zhenhua and Tang, Ruiming and Li, Ruixuan},
  journal={arXiv preprint arXiv:2506.23643},
  year={2025}
}
